# UNISWAP: Impermanent Loss and Risk Profile of a Liquidity Provider

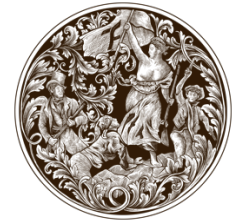

**Andreas A Aigner[*], Gurvinder Dhaliwal[+]**

[*] TradeFlags, Vienna, Austria, [+] Fuel Ventures, London, United Kingdom

Uniswap is a decentralized exchange (DEX) and was first launched on November 2, 2018 on the Ethereum mainnet [1] and is part of an Ecosystem of products in Decentralized Finance (DeFi). It replaces a traditional order book type of trading common on centralized exchanges (CEX) with a deterministic model that swaps currencies (or tokens/assets) along a fixed price function determined by the amount of currencies supplied by the liquidity providers. Liquidity providers can be regarded as investors in the decentralized exchange and earn fixed commissions per trade. They lock up funds in liquidity pools for distinct pairs of currencies allowing market participants to swap them using the fixed price function. Liquidity providers take on market risk as a liquidity provider in exchange for earning commissions on each trade. Here we analyze the risk profile of a liquidity provider and the so called impermanent (unrealized) loss in particular. We provide an improved version of the commonly denoted impermanent loss function for Uniswap v2 on the semi-infinite domain. The differences between Uniswap v2 and v3 are also discussed.

***JEL-*** A10, B10, D40, D47, D53, E44, F30, F60, G10, G14, G21, G23, G51, I10, K10, L14, M10, O16, O31, O33, O40, P10, C63, C70, D83, D85.

## I. Introduction

Decentralized Finance is a unique new application to the world of cryptos [2-41]. An easy introduction to this sector is found in [23] and [20]. It offers products and services akin to centralized finance but instead is decentralized. Meaning that where there was a central authority handling all the transactions, there is none. Market participants interact directly with other participants through the use of predefined contracts, smart contracts. There are also novel products that are unique to decentralized finance. Uniswap a decentralized exchange with automated market making functionality is one example. Uniswap started off with offering exchanges between Ethereum and other currencies (or tokens/assets) [1]. Version 2 generalized this to the exchange between any pair [42]. Liquidity providers of v2 provide liquidity on a semi-infinite domain of the exchange rate, whereas v3 allows liquidity providers to set an arbitrary finite range instead, thereby maximizing the use of their pooled assets [43]. Uniswap uses a fixed price function, also known as a liquidity function, to set the price for each trade. The function is called a market making function and the programmatic use of this function to offer prices is commonly referred to as *automated market making*. It has its origins in prediction markets, online ad auctions and instructor rating markets, where one of the most used scoring rule is the logarithmic market scoring rule (LMSR) [44, 45].

Historically market making has been studied widely [46-61], since it has its origin in the stock and forex market, where firms would take the role of a designated market maker (DMM) on the exchange in return for being able to profit from the liquidity of the market. With the advent of high frequency trading and algorithmic trading such automated means of trading can, in a sense, also be regarded as automated market making. In the derivatives market the option market maker would be quoting bid-offer prices for options on a whole range of stocks listed on an exchange in order to provide liquidity and collect premium and commissions in return [62, 63]. These methods are also referred to as algorithmic market making or quite often called automated market making as well. The deterministic market making functions are normally not used in these situations, instead they are sophisticated stochastic models acting as a feedback loop, responding elastically to market supply and demand.

The market making functions used in decentralized finance are deterministic. Of those that exist today there are several different decentralized exchanges that use different market making functions and a number of papers have been written related to them [44, 45, 64-89]. Some other examples are Bancor, Balancer, Curve, Sushiswap.

The reason why decentralized exchanges have turned to deterministic market making functions is because running a decentralized exchange with an order book is simply unfeasible. The transactions would be slow and it would be very expensive to run. A deterministic function can be easily coded in a smart contract on the Ethereum chain instead of replicating the dynamics of an order book similar to a centralized exchange. The disadvantage of course is that the liquidity across the whole price domain is fixed too at every time, provided there is no in or outflow of liquidity.

But an advantage a decentralized exchange has though is that it is decentralized, permissionless, secure, censorship-resistant and automated, void of any third-party interaction [64]. It operates by various participants interacting directly with each other (through smart contracts). It is secured by the Ethereum network that is designed to be hackproof. It is censorship resistant since participants from anywhere in the world can interact with each other without the need for approval or KYC. It is automated in the sense that, possibly except for the web interface/front end itself, it is extremely unlikely that the smart contracts will stop working. The decentralized exchange will always be available 24/7.

A novel key advantage a decentralized exchange has to offer, is the fact that market participants can earn income on their balances, not without any risks though. But it is important to highlight here that centralized exchanges or platforms such as Coinbase collect enormous transactions fees (Coinbase Revenue 2020 was $1.27bn). Decentralized exchanges such as Uniswap offer market participants a very attractive opportunity to earn such transaction fees. By depositing funds into a liquidity pool, they can earn on every transaction. Market participants who use the exchange facility pay a fixed fee (0.05%, 0.3% or 1%) depending on the liquidity pool and pair they want to trade. These fees are fixed and users know that these fees get paid directly to other market participants being liquidity providers.

Considering the revenue of crypto exchange companies such as Coinbase, this offers up an alternative business model for crypto and a potentially lucrative use case for crypto currencies in general.

Average returns for liquidity providers in the biggest liquidity pool on Uniswap make around 1% per week (52%pa) in return for the last couple of weeks before the time of writing [90]. There are smaller liquidity pools with more illiquid currencies/tokens that can have extreme average returns, but they don't come without any serious risks involved (see recent case of Mark Cuban's Titan-DAI liquidity pool disaster [91, 92]).

Novel investors and curious crypto enthusiasts will want to get a firm grip on understanding exactly what the risks are in being a liquidity provider. Here a detailed analysis of the risk profile is given, the paper is grouped into the following sections: Section II describes the Market Making Function, Section III discusses Impermanent Loss and the Risk Profiles associated with Uniswap v2 and v3 and Section IV concludes with a summary. Appendix I has a derivation of the impermanent loss and Appendix II is a guide on How to become a Liquidity Provider.

## II. MARKET MAKING FUNCTION

The market making function that Uniswap v2 uses is the constant product function and can be described by this equation

$$x \cdot y = K \qquad \text{Eq. (1)}$$

where x and y are the amounts of each asset in the liquidity pool and K represents the total amount of liquidity. Furthermore, it is required that the ratio of the two assets has to always represent the exchange rate of the two assets P.

$$\frac{x}{y} = P \qquad \text{Eq. (2)}$$

The relationship of x and y can be plotted as in Figure 1 where the exchange rate P goes to infinity for X large and Y small and approaches zero for the opposite case. Each point along this curve therefore represents a predefined ratio of amounts x and y making up the liquidity pool.

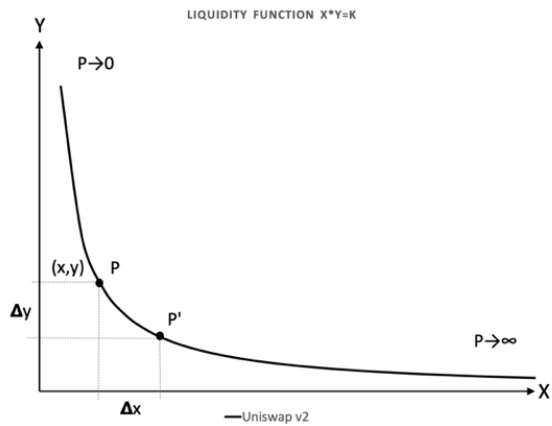

*Figure 1: Liquidity function of Uniswap v2.*

If someone wants to purchase an amount of $\Delta y$ reducing the amount of liquidity in y, he will be quoted an amount of $\Delta x$ that exactly matches the constraint given by the market making function. Both the current Price of the liquidity pool (x,y) and the new price of the liquidity pool (x+$\Delta x$, $y - \Delta y$) have to satisfy Eq. (1) such that the product of both is equal to the constant K. This means that we can solve this equation

$$x \cdot y = (x + \Delta x) \cdot (y - \Delta y)$$

for $\Delta x$

$$\Delta x = \frac{\Delta y}{y - \Delta y} \cdot x \qquad\qquad Eq.\ (3)$$

The amount of $\Delta x$ charged for every amount $\Delta y$ offered is therefore known and defined along the whole liquidity curve. We can notice that the more asset x is depleted the higher the amount of $\Delta y$ is charged for a fixed amount of $\Delta x$ and likewise the more asset y is depleted the higher the amount of $\Delta x$ is charged for a fixed amount of $\Delta y$.

Furthermore, we note here that solving Eq. (1) and Eq. (2) for x and y we can express the liquidity reserves in terms of the Price P and the constant K as

$$x = \sqrt{K \cdot P} \ \text{ and } y = \sqrt{\frac{K}{P}} \qquad\qquad Eq.\ (4)$$

In Uniswap v3 the liquidity is limited to a range in prices $P \in \left[P_{Low}, P_{High}\right]$ at which one of the liquidity amounts is zero at either end. This means that when the price of the asset reaches the lower bound then all the assets are in asset y and when the price of the asset reaches the upper bound all the assets are in asset x. This amounts to a shift in the liquidity function, Eq. (1), which can be expressed as

$$x' \cdot y' = K \qquad\qquad Eq.\ (5)$$

where

$$x' = x + \sqrt{K \cdot P_{Low}} \qquad \text{and} \qquad y' = y + \sqrt{\frac{K}{P_{High}}} \qquad\qquad Eq.\ (6)$$

The intersections of this liquidity curve with the x and y axis is given by

$$x = 0, \qquad y_{max} = \sqrt{K} \cdot \left(\frac{1}{\sqrt{P_{Low}}} - \frac{1}{\sqrt{P_{High}}}\right) \qquad\qquad Eq.\ (7)$$

and

$$x_{max} = \sqrt{K} \cdot \left(\sqrt{P_{High}} - \sqrt{P_{Low}}\right), \qquad y = 0 \qquad\qquad Eq.\ (8)$$

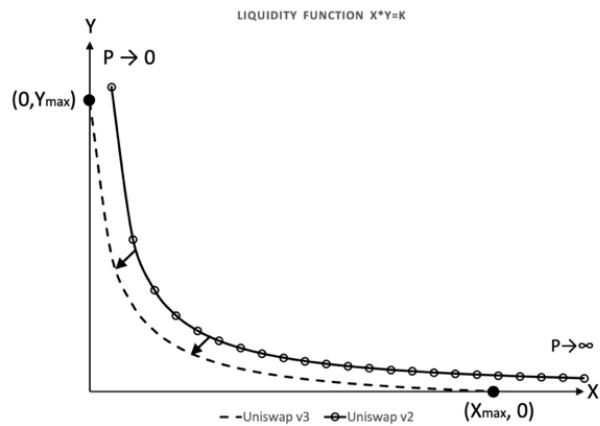

*Figure 2: Liquidity function of Uniswap v3*

corresponding to the points $(0, y_{max})$ and $(x_{max}, 0)$ in Figure 2.

Each Price P in the range $P \in [P_{Low}, P_{High}]$ corresponds to a pair of points $(x, y)$ between $(0, y_{max})$ and $(x_{max}, 0)$. Outside of this range the liquidity pool consists solely of one of the assets x or y.

When we think of an order book market we think of a distribution of orders below and above the current market price (spot price), as is sketched in Figure 3. The distribution of bid and offers might be symmetric or asymmetric. Bid-offers might be concentrated around the spot price and fall off away from it, which is referred to as convex (higher spot liquidity), or increase with prices away from the spot price (low spot liquidity), which is called concave. In the case of an automated market making, such as Uniswap, one can think of the quantities available to trade at different bid offer prices as well and it will look something similar to what is shown in Figure 3. The equivalent order book has increasing amounts of asset y the lower the price of y is with respect to x and less on offer the higher the price of y is. The amounts can be thought of as bucketed and are represented by the amount of $\Delta y$ for each $\Delta x$. This corresponds to the tick sizes in the common stock market. While we might not have a symmetric order book around a current spot level, we do still have an equivalent order book, which is not dynamic though and stays fixed as long as there is no in or outflow of liquidity.

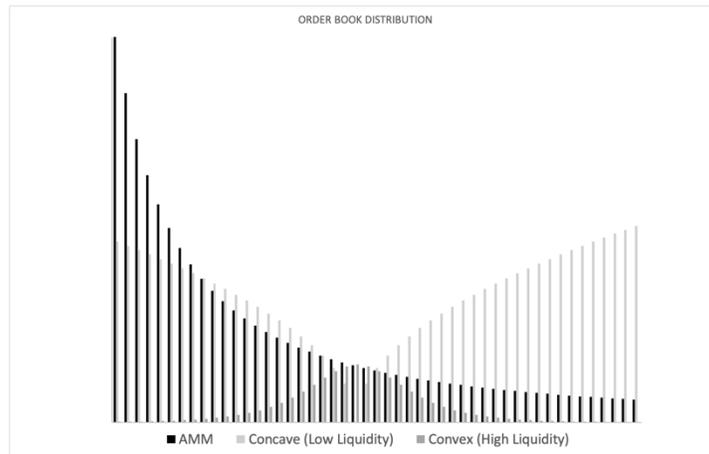

*Figure 3: Examples of order books and market depth.*

## III. IMPERMANENT LOSS & RISK PROFILE

Impermanent loss is the loss incurred by a market making position versus keeping the initially allocated amounts fixed. There are a number of online calculators for the impermanent loss such as [93-96]. The Uniswap website also has a formula for it given here [97] and there are a couple of papers and websites containing the same formula for it here [68, 98, 99]. The formula quoted there is incorrect and will derive the correct impermanent loss equation here. Please note that the term impermanent loss can rather be thought of as what is usually referred to as an unrealized loss. This is a more common way to call such a loss, which only becomes realized if one chooses to sell out of a position which has dropped in value.

Given an initial Price $P_0$ the value $V_0$ of a 2-asset portfolio initially is given by

$$V_0 = x + P_0 \cdot y \qquad \qquad Eq. (9)$$

The new value $V_1$ when traded at a new price $P_1$ is given by

$$V_1 = x' + P_1 \cdot y' \qquad \qquad Eq. (10)$$

Compare this to a 2-asset portfolio with fixed quantities where the value initially would be the same as above, but the value at the new price would also be given by

$$x + P_1 \cdot y \qquad\qquad \textit{Eq. (11)}$$

The impermanent or unrealized loss is the difference between the portfolio change of the market making portfolio and the change in value of a portfolio of assets with fixed quantities. This is the loss on top of a mark to market move of an equivalent fixed-quantity portfolio.

It can be written as

$$x' + P_1 \cdot y' - (x + P_0 \cdot y) - \big(x + P_1 \cdot y - (x + P_0 \cdot y)\big) \qquad\qquad \textit{Eq. (12)}$$

which simplifies to

$$x' - x + P_1 \cdot y' - P_0 \cdot y - (P_1 - P_0) \cdot y \qquad\qquad \textit{Eq. (13)}$$

Substitution and cancelling of terms yield the following equation for the impermanent loss $\varepsilon$

$$\varepsilon = \frac{\Delta PNL}{V_0} = \sqrt{R} - \frac{1}{2} \cdot (R + 1) \qquad\qquad \textit{Eq. (14)}$$

Which is expressed in terms of R, the ratio of the new versus the old price $R = {P_1}/{P_0}$. A detailed derivation with all the steps involved is provided in Appendix I.

When we chart this and compare it to the calculations from the other sources we get a loss which is less pronounced for big drops, but bigger for larger price moves, see Figure 4. When the price drops by 1/2 you have a loss $-4.29\%$ as compared to $-5.72\%$. On the contrary when the price increases by 2X you have a loss of $-8.58\%$ compared to a loss of $-5.72\%$ using the equation commonly quoted. For a 2/3 drop this difference increases to $-9.05\%$ and 13.62%, versus $-26.79\%$ and 13.40%. The reason being is that the impermanent loss equation commonly quoted is calculated relative to the final value of the portfolio. Here we calculate it versus the initial value of the portfolio which makes more sense as an investor. Since looking at a PNL relative to the final PNL of a fixed portfolio is misleading. Investors will always be surprised that their upside loss was higher than expected and their downside loss was lower. It is particularly misleading when you are trying to compare different strategies with each other, since the final value of the portfolio will be different for each strategy.

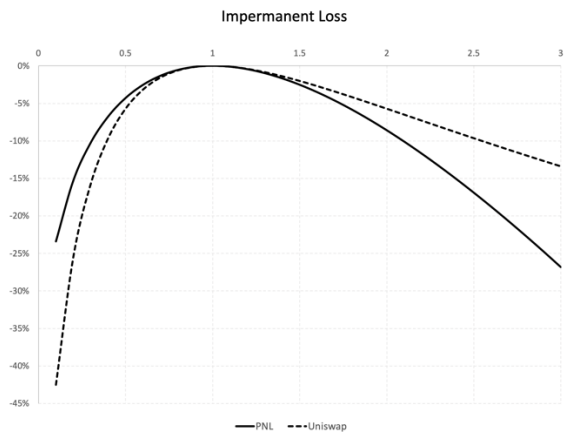

*Figure 4: Impermanent Loss*

This is the impermanent loss on top of the mark to market of the initial position in assets x and y. If we want to know the total PNL we have to include the mark to market of the initial position. Figure 5 shows a normalized version compared to a buy and hold with fixed position sizes. A buy and hold portfolio has a linear relationship with its value since the value of the portfolio increases or decreases with price linearly. The liquidity provider has a near linear relationship with price moves around the initial price only. When the price drops to zero the portfolio value drops sharply to zero, since the liquidity provider is buying more and more of the asset losing its value until the price is exactly zero and all of the other liquidity asset has dropped to zero. When the price increases the liquidity provider loses more and more of the upside gains since he is selling more and more of the appreciating asset as the price increases. This is the reason why the liquidity provider is losing money in both directions. He is buying the asset that is dropping in value and selling the asset that is rising in value. Ideally the liquidity provider wants the asset prices to hover around his initial price, so one can therefore think of the liquidity provider being 'short volatility' and 'short convexity'. His position looks similar but is not identical to someone being short a call and put option (a straddle). The risk profile here is nonlinear.

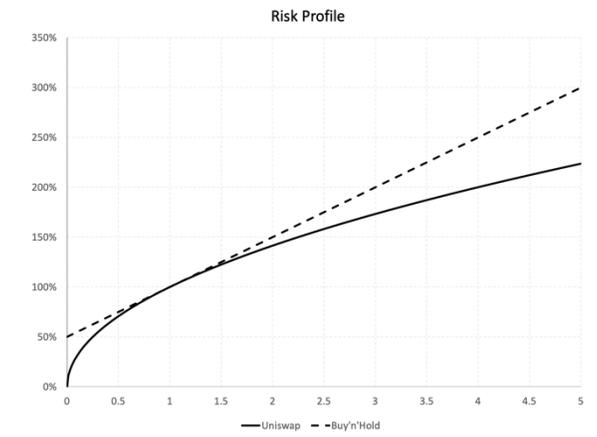

*Figure 5: Risk Profile of a Uniswap v2 liquidity provider and a Buy'n'Hold*

The impermanent loss of Uniswap v3 can be calculated similarly to v2, first one needs to solve Eq. (5) and Eq. (6) for x and y, similarly to Eq. (1),Eq. (2) and Eq. (4), which amounts to finding the roots of a quadratic equation in x and y.

$$x^2 + \sqrt{K}\left(P\frac{1}{\sqrt{P_{High}}} + \sqrt{P_{Low}}\right) \cdot x + KP \cdot \left(\frac{\sqrt{P_{Low}}}{\sqrt{P_{High}}} - 1\right) = 0$$

and

$$y^2 + \sqrt{K}\left(\frac{1}{\sqrt{P_{High}}} + \frac{\sqrt{P_{Low}}}{P}\right) \cdot y + \frac{K}{P} \cdot \left(\frac{\sqrt{P_{Low}}}{\sqrt{P_{High}}} - 1\right) = 0$$

Which can be solved using the ordinary p-q formula. Substituting this in the formula for the difference in the market to market move of a 2-asset market making portfolio compared to a fixed quantity portfolio yields the impermanent loss of a Uniswap v3 liquidity provider. We note from these quadratic equations that when $P_{Low}$ is zero and $P_{High}$ is infinite we return the same equations as given Eq. (4) in.

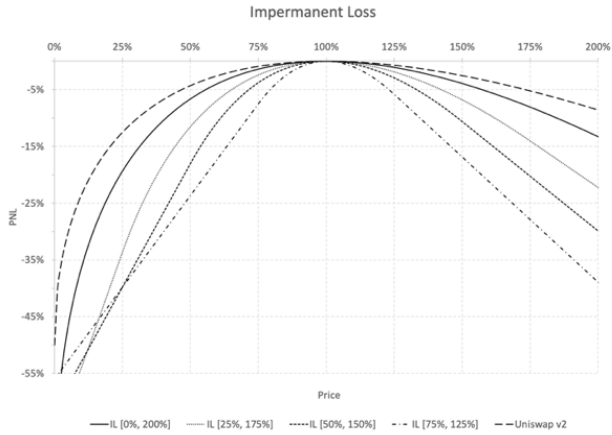

*Figure 6: Impermanent Loss of various Uniswap v3 liquidity positions and a Uniswap v2 liquidity position*

Examples for various price ranges are shown in Figure 6 and compared to the impermanent loss of a Uniswap v2 liquidity provider. It is obvious that when the fixed range of the liquidity provider approaches the semi-infinite domain of the v2 liquidity provider the impermanent loss functions become similar. For smaller ranges the impermanent loss gets more symmetric and decreases around the initial value having steeper losses than an ordinary Uniswap v2 position with a semi-infinite domain will have.

For example, when the price moves by 20% the impermanent loss of a v2 liquidity position will be −0.56% and −0.46% while a fixed range of 25% and 125% of initial will have an impermanent loss of −4.75% and −3.8%. Some more examples are collected in Table 1.

*Table 1: Impermanent Loss of various Liquidity Positions*

| %Move/Range | -20% | Initial | 20% |
|---|---|---|---|
| [0%, inf) | -0.56% | 0 | -0.46% |
| [0%, 200%] | -0.86% | 0 | -0.70% |
| [25%, 175%] | -1.5% | 0 | -1.22% |
| [50%, 150%] | -2.34% | 0 | -1.91% |
| [75%, 125%] | -4.75% | 0 | -3.8% |

Consider a fixed range of 80% and 120%, the risk profile of such a liquidity position will look like Figure 7. Since the liquidity outside of this range is composed entirely of one or the other asset the value of the portfolio will decline linearly below the lower bound, since the pool consists of 100% of the declining asset, and remain constant above the upper bound, since the pool consists of 100% of the asset that is not appreciating. Compare this to a buy and hold portfolio of the same initial asset distribution, which increases/decreases linearly with price. Note that when the price drops to zero for one asset your buy and hold portfolio will still be worth 50% whereas the market making portfolio will be have lost 100% of its value since it was long 100% of the declining asset already since it breached the lower bound.

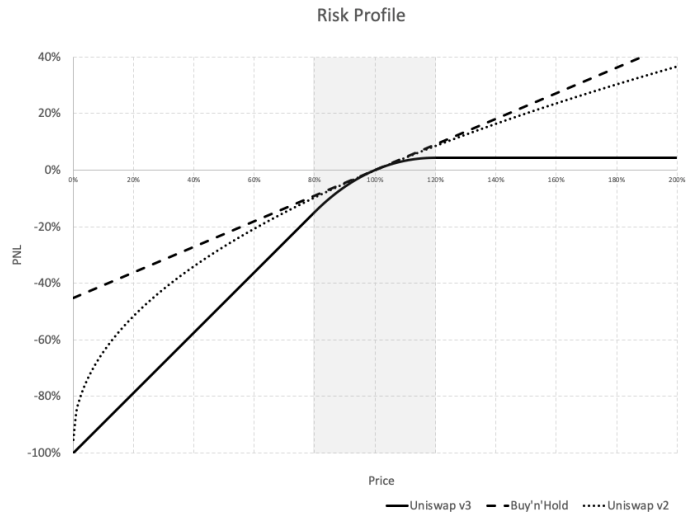

*Figure 7: Risk Profile of an 80%-120% liquidity position.*

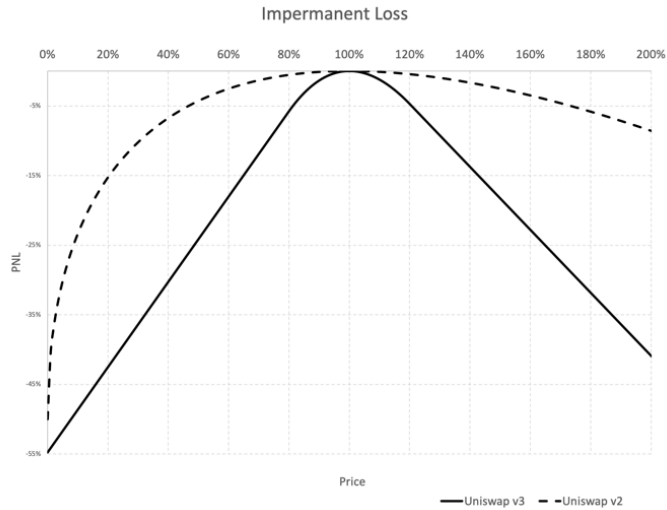

*Figure 8: Impermanent loss of an 80%-120% liquidity position.*

Figure 8 shows the corresponding impermanent loss of a [80%, 120%] range liquidity provider. We note here that a liquidity provider portfolio risk profile is always at a disadvantage to a buy and hold portfolio. Some websites interactive calculators erroneously show a risk profile that indicate that you will lose less than a buy and hold portfolio. This is not the case. In essence as exhibited by the PNL profile, Figure 7 and Figure 8, it is similar to being synthetically short variance (short Gamma) on the underlying, earning Theta (yield). Extreme spot price moves will manifest this short gamma; in further papers we will discuss periodic delta hedging strategies and short dated puts to ameliorate and subdue the risk characteristics endemic in the form of liquidity positions discussed in this paper.

## IV. CONCLUSION

Current investors can purchase crypto using centralized exchanges (CEX), which are companies in the conventional sense that provide a platform or app to deposit fiat currencies into or offer the option to purchase using credit cards. They provide an on-ramp facility for investors to exchange fiat currencies into crypto. Many of these platforms offer interest rates on crypto (and fiat) deposits and further investment products such as locked staking.

In contrast to centralized exchanges there are decentralized exchanges (DEX) which have no central entity managing the deposits. Instead the interactions between participants are handled by smart contracts on the Ethereum chain. Assets get exchanged using a smart contract using a deterministic market making function that has a set price for every amount of token that gets offered or bid. This function of liquidity represents a distribution of bid offers similar to order book type markets provided by centralized exchanges. Instead of having a dynamic order book with market depth the depth of bid-offers is fixed, as long as the amount of liquidity underlying doesn't change. The liquidity in such a liquidity pool is not provided by a centralized entity, as would be the case on a CEX, but instead by other individual market participants who get compensated by trade commissions in return for the risk they take on. The risk that a liquidity provider takes on is essentially of two kinds. When setting up a position initially an amount of each asset is provided and usually in a ratio of about 50% each. A liquidity provider therefore has, without any other interaction, a linear risk of the price of one of the assets changing versus the other asset. This is normally called Delta risk, since Delta in the option market is the rate of change of one asset with the change in the other asset. Usually the other asset in such case is the base currency such as the US Dollar. When you have a liquidity position you can have two assets which are both different to your US Dollar, so in practice you have Delta risk on both currencies already.

The second risk that a Liquidity Provider has is the change in his position due to other market participants interacting with his liquidity pool. Every time part of his position gets bought or sold, the price of the asset changes. And at either one of the extreme ends of the prices, albeit 0 and infinity or a fixed range ($P_{Low}$ and $P_{High}$) he will have swapped one of the assets into the other assets completely. The difference between the value of the portfolio of two assets with and without these transactions is called impermanent loss. It can also be called unrealized loss in this situation, because if the price of the asset reverses to the initial price a liquidity provider would end up with exactly the same position as what he set out with, having zero loss, but would have earned commissions along the whole price swing. The Delta risk above is also an unrealized loss, since the Liquidity provider ends up with zero loss or gain when the price drops back to the same initial price.

To summarize both the Delta as well as the impermanent loss are unrealized as long as the liquidity provider does not withdraw his liquidity from the pool. At the point where the liquidity provider withdraws his funds from the pool, is when his loss or gain due to the impermanent loss gets realized. His loss or gain due to the Delta risk is only realized once he sells out of his position or swaps one of the assets for the other.

Here we have described the underlying mechanics of a market making function in Section II, and have derived the impermanent loss function for Uniswap v2 and v3, in Section III. We have provided an improved impermanent loss formula for the commonly quoted Equation for v2 and online calculators for v3. We have charted the risk profile of positions in v2 and v3 and compared various different ranges of liquidity, showing that v2 is approached in the limit of the range going to infinity.

Uniswap is one of the most liquid decentralized exchanges. There are other exchanges that offer similar market making products that will work according to different market making functions. There is still ongoing research into these which can be found in our references below.

What other risks exist for a liquidity provider? One of the key properties of a liquidity provider is that he is willing to own all of either one of the assets at either his lower price limit or zero (in the case of v2). This means that when one of the assets has huge price swings or is compromised, as for example in the recent case of Mark Cuban (ironfinance, Titan-DAI). The liquidity provider would have sold all of his reliable assets to purchase the compromised asset. A liquidity provider cannot be certain that he will own any or part of either asset at the time of redemption. If a liquidity provider is happy to own either one of the assets at its low and sell out of it at a higher level, he will get compensated for this through the commissions he will earn when participants are trading on his liquidity pool.

Since market making functions are using individual liquidity curves there are also various arbitrage opportunities. Some with external order book type exchanges as well as internally between a trio of currency pairs that end up quoting different cross exchange rates for the exchange rate in the first pair [72]. Provided the arbitrage opportunity is greater than the transaction costs required it will be utilized. The work involved in taking advantage of these opportunities is extensive, especially if its across different exchanges. There is however the advantage of using atomic transactions, meaning one smart contract transaction that will exercise a round trip in the arbitrage. Building such arbitrage tools as a regular person will be too difficult, but suffice to say it is most certainly done already to some extent. In fact decentralized exchanges such as Uniswap can represent a 'fair' price of assets since it will have been arbed already, these are called price oracles and Uniswap is regarded as such as well [44] .

As of time of writing the biggest liquidity pool on Uniswap (ETH-USDC) is earning an average of 1.5% weekly [90, 100, 101], which amounts to 78% annually at the current trading volumes. A liquidity provider who provides an equal amount of liquidity mapped along the price curve can expect to earn the same or similar return.

There are other risks involved with crypto which have to do with smart contract risk, fraudulent DEXs, regulatory clampdowns, ISP provider censorship and other exterior risks [102-106]. There are discussed widely in the literature here and here. Useful websites for monitoring the Defi space and security issues are vfat.tools, RugDoctor and rekt.news.

APPENDIX I: DERIVATION OF THE IMPERMANENT LOSS FUNCTION V2

We start off with the change in the Value of a portfolio of a liquidity provider versus the change of portfolio of a fixed-asset portfolio.

$$V_1 - V_0 - (V_{fixed} - V_0)$$

Notice how we can cancel the $V_0$ term and write

$$x' + P_1 \cdot y' - (x + P_1 \cdot y)$$

Substituting Equations 1,2 in Equation x gives

$$\sqrt{K \cdot P_1} + P_1 \cdot \sqrt{\frac{K}{P_1}} - \left( \sqrt{K \cdot P_0} + P_1 \cdot \sqrt{\frac{K}{P_0}} \right)$$

When you expand this you get

$$\sqrt{K} \cdot \left( \sqrt{P_1} - \sqrt{P_0} \right) + \sqrt{K} P_1 \cdot \left( \sqrt{\frac{1}{P_1}} - \sqrt{\frac{1}{P_0}} \right)$$

The initial value of the portfolio is

$$V_0 = x + P_0 \cdot y = \sqrt{K \cdot P_0} + P_0 \cdot \sqrt{\frac{K}{P_0}} = 2\sqrt{K \cdot P_0}$$

Dividing this by $V_0$ we get

$$\frac{\sqrt{K} \cdot \left( \sqrt{P_1} - \sqrt{P_0} \right)}{2\sqrt{K \cdot P_0}} + \frac{\sqrt{K} P_1}{2\sqrt{K \cdot P_0}} \cdot \left( \sqrt{\frac{1}{P_1}} - \sqrt{\frac{1}{P_0}} \right)$$

Cancelling out the terms gives

$$\frac{\left( \sqrt{P_1} - \sqrt{P_0} \right)}{2\sqrt{P_0}} + \frac{P_1}{2\sqrt{P_0}} \cdot \left( \frac{\sqrt{P_0} - \sqrt{P_1}}{\sqrt{P_0} \cdot \sqrt{P_1}} \right) = \frac{1}{2} \left( \sqrt{\frac{P_1}{P_0}} - 1 \right) + \frac{1}{2} \sqrt{\frac{P_1}{P_0}} \cdot \left( 1 - \sqrt{\frac{P_1}{P_0}} \right)$$

When we introduce the ratio of prices

$$R = \frac{P_1}{P_0}$$

This can simply be written as

$$\varepsilon = \frac{1}{2} \left( \sqrt{R} - 1 \right) + \frac{1}{2} \sqrt{R} (1 - \sqrt{R})$$

Which becomes

$$\varepsilon = \sqrt{R} - \frac{1}{2} \cdot (R + 1)$$

Comparing this to the commonly quoted impermanent loss function, where the change in value as relative to the final portfolio value is calculated instead of the initial value, you would divide by $V_{fixed}$ instead of $V_0$ and get [97, 107-109]

$$\varepsilon = \frac{2\sqrt{R}}{1+R} - 1$$

APPENDIX II: HOW TO BE A LIQUIDITY PROVIDER

We want to explain the steps involved and the fees involved in initiating a liquidity position. We choose the most liquid and biggest liquidity pool on Uniswap which is ETH vs USDC. USDC is a stablecoin which is backed by US Dollar 1-1. The company which issues these stablecoins is backed by investors such as Coinbase.

1. First, we transfer by wire transfer or SEPA say 2000 USD to a centralized exchange for free.
2. Then we buy 1000$ worth of ETH and incur a fee of say 0.00061814 ETH ($1.19 with ETHUSD = 1930$).
3. We transfer this ETH to a wallet accepted by Uniswap and incur a fee of 6.98$ (0.0036ETH). We use Metamask.
4. We purchase 1000$ worth of USDC for a fee of about $1.19.
5. We also transfer this amount of USDC to the accepted wallet and incur a fee of 8$
6. Connecting your wallet to Uniswap we find the most liquid pool using the charts. One of the options is to provide liquidity. Since all the liquidity pools have to use a wrapped ETH (WETH9) to provide liquidity we have to convert the ETH we have in our wallet to WETH9. We do this using Uniswap which incurs a 1.17$ fee.
7. Now we can start with setting up the liquidity position. We choose to allocate all of the wrapped ETH (WETH9) and it will return us the required USDC we need to allocate. If the balance of ETH versus USDC is not exactly 50:50 then you will need to transfer for USDC from as described above. Alternatively start off with buying more of the assets or reducing the amount of ETH to deposit.
8. In our example we end up having a balance of 58% versus 42% roughly.

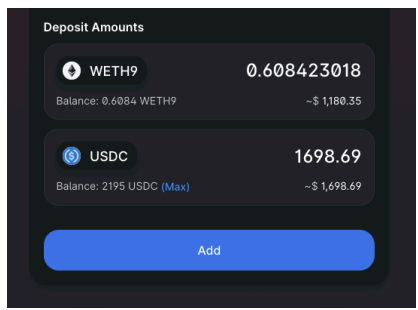

9. Then one confirms the wrapped ETH and USDC position which costs 1.41$ and 1.87$ in fees.
10. After waiting for these two transactions to go through you can confirm the liquidity position as a whole, which costs 13.26$ to mint.

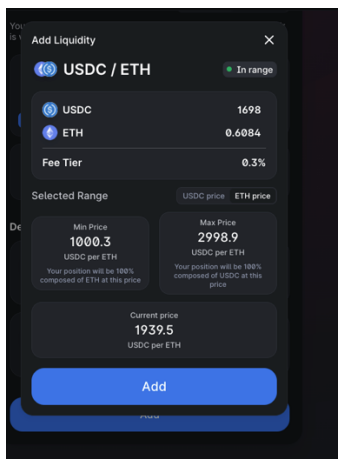

11. After paying this last fee after a couple of minutes have passed one will see the position listed under 'your positions' and can

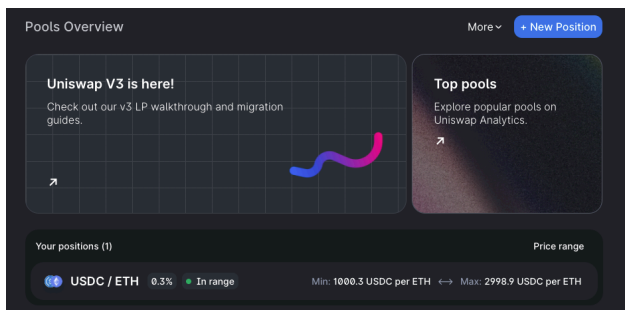

12. View the Ethereum Address of this position on the Ethereum chain using Etherscan. One can also view the running fees collected for the liquidity pool. Which in this case is already around 0.54$ in fees after about 1-2hrs online.

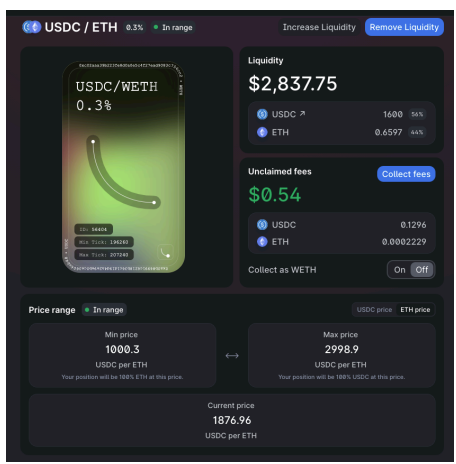

In the whole process we have spent 35,07 USD or in this example 1.25% of the total value of the portfolio, which is less than the theoretical return in commissions in one week (1.5%).

AUTHORS

**Email:**      Andreas A. Aigner, andreas@tradeflags.com
               Gurvinder Dhaliwal, guv@fuel.ventures